\documentclass[prb,aps,twocolumn]{revtex4}
\usepackage{graphicx}

\begin{document}

\renewcommand{\thefootnote}{\fnsymbol{footnote}}
\draft
\title{Cavity polariton optomechanics: Polariton path to fully resonant dispersive coupling in optomechanical resonators}

\author{G. Rozas$^{1}$, A. E. Bruchhausen$^{1}$, A. Fainstein$^{1, }$\footnote{email:afains@cab.cnea.gov.ar}, B. Jusserand$^2$, and A. Lema\^itre$^3$}
\affiliation{$^1$Centro At\'omico Bariloche \& Instituto Balseiro, C.N.E.A.,8400 S. C. de Bariloche, R. N., Argentina}
\affiliation{$^2$Institut des NanoSciences de Paris, UMR 7588 C.N.R.S. - Universit\'e Pierre et Marie Curie, 75015 Paris, France}
\affiliation{$^3$Laboratoire de Photonique et de Nanostructures, C.N.R.S., 91460 Marcoussis, France}

\begin{abstract}
Resonant photoelastic coupling in semiconductor nanostructures opens new perspectives for strongly enhanced light-sound interaction in
optomechanical resonators. One potential problem, however, is the reduction of the cavity Q-factor induced by dissipation when the resonance is 
approached. We show in this letter that cavity-polariton mediation in the light-matter process overcomes this limitation
allowing for a strongly enhanced  photon-phonon coupling without significant lifetime reduction in the strongly-coupled regime. Huge optomechanical coupling factors in the PetaHz/nm range are envisaged, three orders of magnitude larger than the backaction produced by the mechanical displacement of the cavity mirrors.

\end{abstract}

\maketitle

Optomechanical resonators, that is, cavities that confine light and mechanical vibrations in the same 
space, strongly coupling the electromagnetic and elastic deformation fields, have emerged as novel 
paradigms for new fundamental ideas and applications.~\cite{Arcizet,Kippenberg-Vahala,Thompson,Eichenfield,Russel,Favero,Marquardt,Groeblacher,Schliesser,Weis,Lin} 
Optomechanical non-linearities, laser cooling, and phonon lasing~\cite{Zhao,Grudinin} have been demonstrated.
In addition, optomechanical devices have been cooled down to the quantum ground state of mechanical motion, signaling a new era of quantum phononics with implications for quantum information processing, sensitive measurements and fundamental research.\cite{O'Connell,Teufel,Chan,Verhagen} 
Very recently hybrid systems combining cavity quantum electrodynamics (CQED) and cavity optomechanics have been theoretically proposed as a means to evidence unconventional dissipative couplings and cooling at the single-polariton level.\cite{Restrepo,Kyriienko}  Here we experimentally demonstrate an additional relevant characteristic of cavity polariton optomechanics, i.e., the possibility to access a hugely enhanced optomechanical coupling of dispersive photoelastic resonant nature, without significant dissipation-induced cavity Q-factor quenching.

Two issues that have been identified as relevant for the development of cavity optomechanics are, on one side, the push 
for ever higher frequencies~\cite{Carmon,Ding,FainsteinPRL2013} and, on the other side, the search for new stronger optomechanical coupling mechanisms.~\cite{Marquardt2,Agarwal,Groblacher,FainsteinPRL2013} 
While micromechanical devices typically oscillate in the KHz-MHz range, GHz-THz frequencies have been attained using nano-size
toroids\cite{Ding} and distributed Bragg reflector (DBR)-based microcavities.~~\cite{FainsteinPRL2013} Radiation pressure is usually identified at the origin of optomechanical coupling. Direct transfer of impulse from the photon field to the resonator mirrors induces vibrations on the latter, which in turn results in a backaction on the electromagnetic field due to the resonator optical detuning induced by the 
mechanical displacement of the mirrors. We have recently reported that GaAs DBR-based microcavities constitute 
optimized optomechanical resonators operating in the GHz-THz range, with the potential of adding an additional photoelastic term to the above described  purely ``mechanical'' mechanism.~\cite{FainsteinPRL2013} The two complementary sides of the coin in this case are electrostriction (for the generation of phonons by light), and the phonon induced modulation of
the dielectric function, i.e., the so-called deformation-potential mechanism (for the backaction of the strain field on the
optical cavity resonance). The photoelastic constant is defined through the relation $\Delta\epsilon$= $P s$.  Here $s$ is the strain associated to the involved vibration, $P$ is the photoelastic coefficient,  and $\epsilon$  is the dielectric constant. While quantitative determinations of photoelastic constants are only available for few materials and only on limited laser wavelength ranges,~\cite{Feldman1968,Jusserand2012,Ultrasonics}  it is understood that under resonant conditions the photoelastic mechanism in GaAs (and similar materials) should become dominant, allowing for huge optomechanical coupling 
factors (in the tens of THz/nm range).~\cite{FainsteinPRL2013}

\begin{figure} 
    \begin{center}
    \includegraphics[scale=0.3,angle=0]{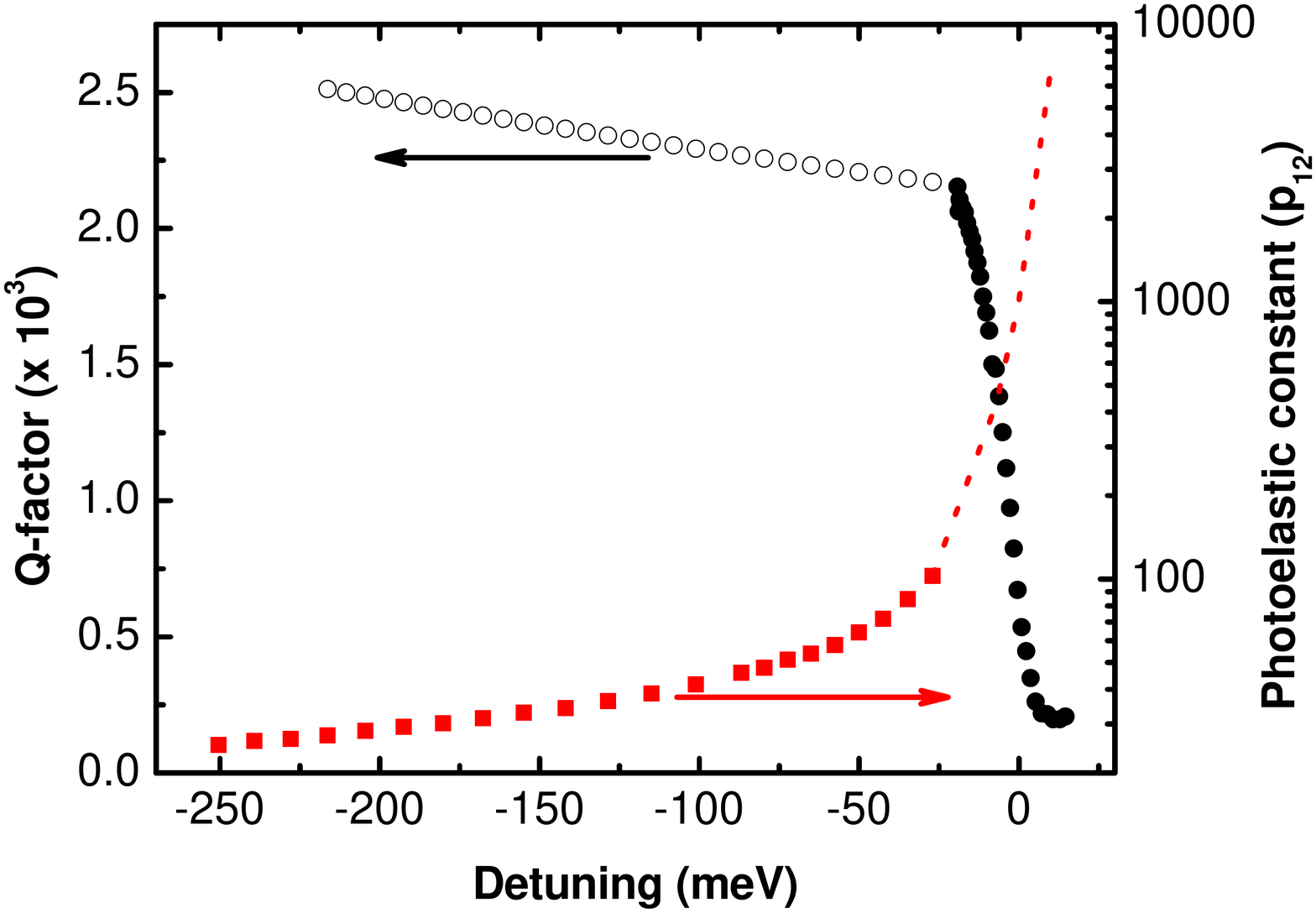}
    \end{center}
    \vspace{-0.8 cm}
\caption{Right scale: Photoelastic constant as a function of the detuning between the laser energy and the $E_0$ gap of GaAs (from Refs.~\cite{Feldman1968,Ultrasonics}). Left scale: optical cavity Q-factor for a $\lambda/2$ GaAs resonator. See text for details.
\label{fig01}}
\end{figure}

An obvious potential limitation for the full use of photoelastic resonant coupling in optomechanical resonators cannot be, however, overlooked.
A resonant enhancement of dispersive photoelastic coupling is always Kramers-Kronig related to dissipation. It is at the electronic gap
that the photoelastic constants resonate. But it is also at the gap where strong absorption sets-in. Thus, it can be expected that at the same pace that photons and phonons increase their coupling due to resonance, the optical cavity Q-factor should be quenched. Indeed this is critical for bulk-GaAs cavities, as is schematically illustrated in Fig.~\ref{fig01}. Here we show together the photoleastic constant of GaAs, and the optical Q-factor of a bulk-GaAs $\lambda/2$ microcavity close to resonance with the fundamental $E_0$ gap. Symbols in the photoelastic constant correspond to data derived from piezo-biregringence experiments,~\cite{Feldman1968} and the dotted curve is a schematic behavior derived from recent predictions based on Brillouin scattering data.~\cite{Jusserand2012,Ultrasonics} The solid symbols in the Q-factor  correspond to measured values, while open symbols are extrapolated assuming no additional mode-broadening in the transparency region of GaAs. Note that due to the $\sqrt{(E-E_0)}$ dependence on energy $E$ of the electron-hole joint density of states in bulk, added to the unavoidable presence of defect and phonon induced absorption below the gap in these materials, absorptions are broad. 
In bulk GaAs and at room temperature excitons are not stable and photons absorbed through the creation of electron-hole pairs are irreversibly lost from the electromagnetic field, resulting in the observed strong reduction of cavity Q-factor as dissipation sets-in. Fig.~\ref{fig01} highlights the fact that there is an enormous potential to exploit photoelastic coupling in resonant materials in the domain of cavity optomechanics, but that in order to do that  dissipative Q-factor quenching needs to be avoided.


Excitons in quantum wells (QWs) could be envisaged as an alternative to circunvent the above limitation because in an absorption process the broad continuum of electron-hole pairs is replaced by a one-to-one correspondence. Excitons define the optical properties of QWs because of the larger binding energy and  oscillator strength induced by confinement.~\cite{Gurioli}  However, in real samples absorption can  be broad and lossy anyway due to roughness and inhomogenoeus broadening (layer thickness fluctuations). This leads for microcavities with embedded collections of QWs in the ``weak coupling regime'' to a general behavior similar to the one displayed in Fig.~\ref{fig01}. As we argue here, the situation is qualitatively modified when we consider the ``strong coupling regime''.~\cite{Polaritons2,Polaritons3,Houdre1} 
The strong coupling limit, as oposed to the weak coupling one, is defined as the regime where the exciton-cavity field interaction evolves faster than the decay rate due to photon cavity or exciton losses. 

Cavity-polaritons, the quasi-particles that define the strong coupling regime, provide the solution to the posed problem essentialy for the following reasons. Firstly, the presence of coupled photon-exciton states allow for extended lifetimes because the dressed states bounce back and forth from exciton to photonic character before the excitation can dephase through the more dissipative channel.~\cite{Sermage} Secondly, polariton mixing leads to much smaller masses, and consequently quantum mechanical effects cause spatial averaging over the disorder potential and hence motional narrowing of the spectral lines.~\cite{Whittaker} And thirdly and most important, inhomogenous broadening becomes irrelevant in the strong coupling regime. In fact a {\em single} linear combination of the set of exciton states couples to the electromagnetic field, and is red-shifted into the transparency region of the material. The remaining $N-1$ remain as uncoupled optically ``dark" states.~\cite{Houdre2,Bongiovanni} The consequence is that polariton states are characterized by the {\em homogeneous} broadening, which in high quality samples can result in Q-factors in the 10$^4$ range.

We will evaluate the optomechanical coupling of a resonator by studying the efficiency for the optical generation of hypersound under polariton excitation in an optical microcavity.~\cite{Alex97} This efficiency is manifested as the Raman cross section for scattering by acoustic phonons.~\cite{SuppInfo}
The sample is a planar microcavity based on GaAs and AlAs materials, with $Al_{0.3}Ga_{0.7}As/AlAs$ ($61.17 / 71.21$~nm) DBRs, 20 pairs in the bottom, 16 pairs on top.  To simplify the Raman experiments the frequency of the mechanical vibrations has been pushed up from the 20 GHz of the basic optomechanical mode of the microcavity as a whole,~\cite{FainsteinPRL2013}  to 250~GHz by replacing the bulk GaAs spacer of the optical microcavity by an acoustic multilayer.~\cite{Trigo,Rozas} The latter mostly consists of a 32 period GaAs/AlAs (14.16/5.81~nm) multiple quantum well (MQW). 
The experiments reported here have been performed in strong resonance with the GaAs exciton of these MQW at 1.53~eV. The optical microcavity has a thickness gradient that allows for the tuning of the cavity mode in the vicinity of this energy.

\begin{figure} 
    \begin{center}
    \includegraphics[scale=0.8,angle=0]{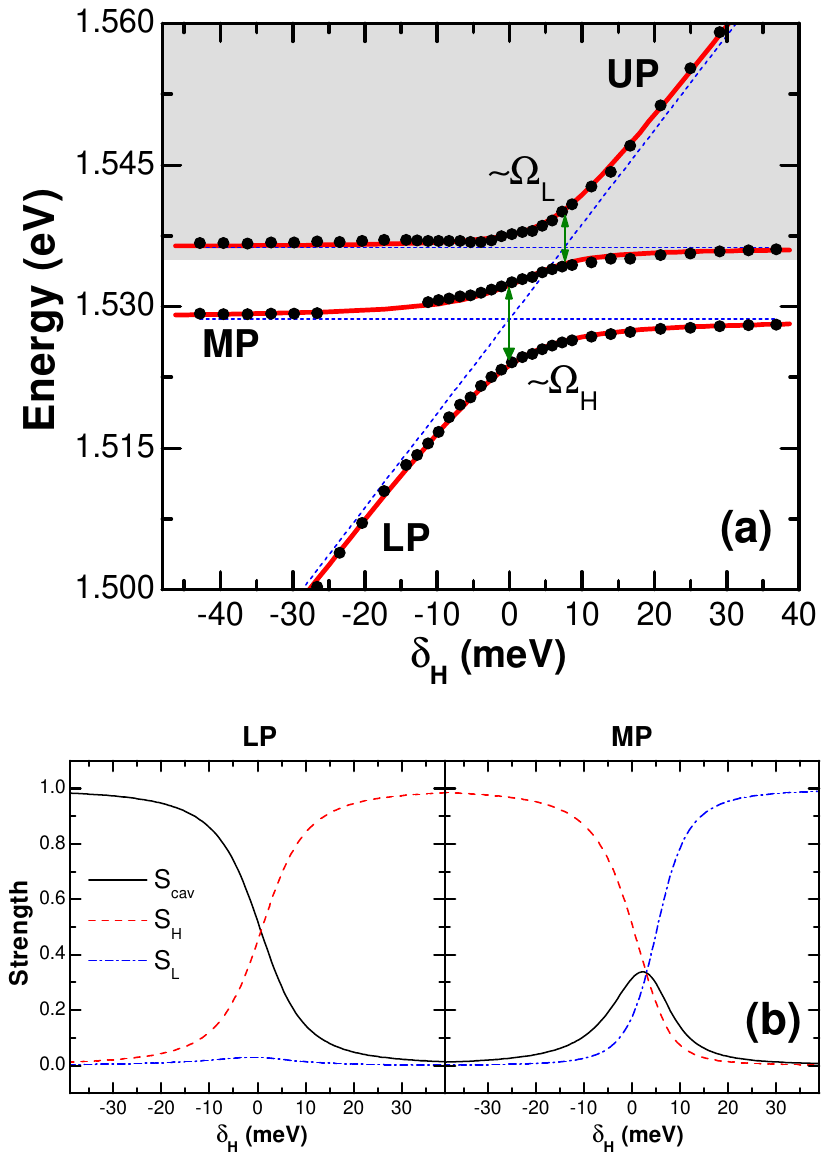}
    \end{center}
    \vspace{-0.8 cm}
\caption{Top panel: Polariton energies (solid symbols) and fitted dispersion (continuous curves) as a function of cavity-$E_1H_1$ detuning ($\delta_H$). LP, MP and UP indicate the lower, middle, and upper polariton branches, respectively. $\Omega_H \approx 9.1$~meV and $\Omega_L \approx 5.9$~meV are the heavy and light-hole Raby splitings, respectively, derived from the fit. The uncoupled
states ($E_1H_1$, $E_1L_1$, and cavity mode) are indicated with dotted curves as a reference.  The grey background marks the continuum of e-h excitations. Bottom panels: Photonic and excitonic strength as a function of cavity-$E_1H_1$ detuning, for the LP and MP polariton branches, derived from the fit shown in the top panel. \label{fig02}}
\end{figure}

Fig.~\ref{fig02}(top) shows with solid symbols the measured polariton energies as a function of cavity-$E_1H_1$ exciton detuning $\delta_H$,~\cite{SuppInfo} together with a simultaneous fit of all three polariton branches, using a three coupled state model which leads to eigenstates of the form:~\cite{Polaritons2,Polaritons3,Houdre1} 
\begin{equation}
\left|P\right>=A_{cav}^{P}(\delta_H)\left|CAV\right> + A_H^{P}(\delta_H)\left|E_1H_1\right> + A_L^{P}(\delta_H)\left|E_1L_1\right>.
\label{eq1}
\end{equation}
The coupling of the exciton states ($E_1H_1$ and $E_1L_1$) derived from heavy ($H$) and light ($L$) holes and the optical cavity mode leads to a three-mode behavior. 
The coefficients $S_{cav}^{P}=\left|A_{cav}^{P}\right|^2$, $S_H^{P}=\left|A_H^{P}\right|^2$, and $S_L^{P}=\left|A_L^{P}\right|^2$ obtained from the fit (and shown in Fig.~\ref{fig02}(bottom) for the lower and middle polaritons) describe the strength of the cavity and excitonic states, respectively, on the dressed polariton states. $P$ stands in general for LP, MP and UP, indicating the lower, middle, and upper polariton branches, respectively.

\begin{figure} 
    \begin{center}
    \includegraphics[scale=1,angle=0]{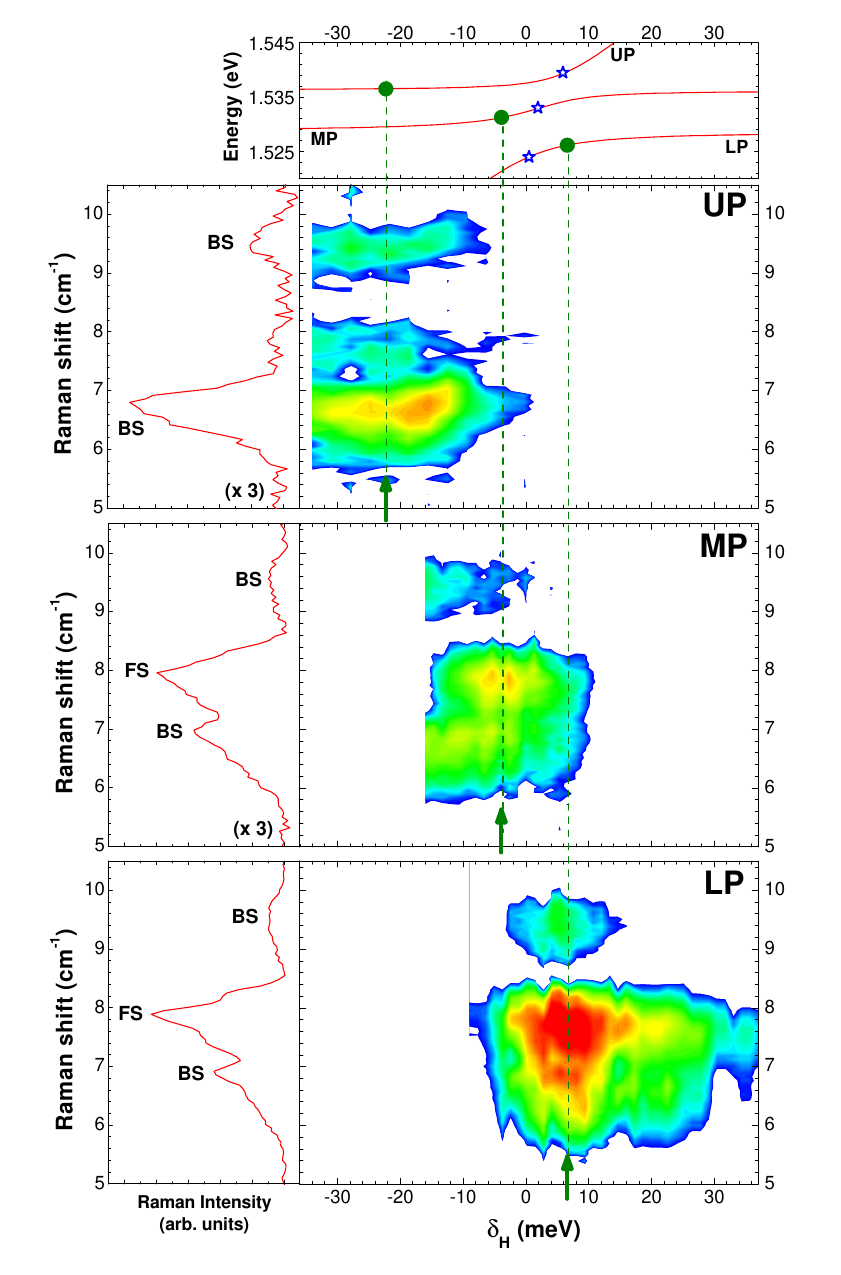}
    \end{center}
    \vspace{-0.8 cm}
\caption{Experimental Raman intensity maps as a function of cavity-$E_1H_1$ detuning $\delta_H$, for the three polariton branches. The color scale is logaritmic, with red the most intense Raman cross section. The top panel shows the calculated polariton dispersion relation. Open stars indicate the expected condition for maximum Raman efficiency, without account of lifetime effects. Left panels present the Raman spectra obtained at the maximum for each polariton branch, labeled with a vertical arrow in the corresponding map (maximum of the integrated intensity including the full Raman spectra).  FS and BS identify modes that are normally seen using forward and back-scattering configurations, respectively.~\cite{Alex95,Rozas}
\label{fig03}}
\end{figure}

Fig.~\ref{fig03} summarizes the main results of this paper. Experimental Raman intensities as a function of $\delta_H$ are displayed as color maps for a double resonant configuration with the laser and collected photons tuned along each of the three polariton branches.
Double resonances are attained using angle tuning.~\cite{Alex95,AlexPRB98}. 
Experiments were performed at 80~K in cuasi-double resonance, with light collected normal to the sample, and the laser incident with $\approx 5^o$. The detuning is varied by shifting the spot position on the sample and adjusting the laser energy to the corresponding local cavity energy. Typical Raman spectra are shown at the left panels for the maximum intensity (integrated over the full Raman spectra) measured on each polariton branch, labeled with a vertical arrow in the corresponding map.  FS and BS identify folded acoustic modes with energies around 8~cm$^{-1}$ (or 250GHz) that are normally seen in Raman experiments using forward and back-scattering configurations, respectively. Both types of modes are simultaneously observed in microcavities due to the standing wave character of the confined electromagnetic field.~\cite{Alex95,Rozas}  As evidenced from these data, the larger intensity of the Raman peaks concentrates close to the branch's anticrossing, that is, when polaritons display similar excitonic and photonic character.

To proceed further in a more quantitative description of the experimental observations we present next the so called ``factorization model'' of Raman scattering mediated by polaritons.~\cite{SuppInfo} Using Fermi's Golden Rule one can obtain:~\cite{Alex97,Bendow,Matsushita,WeisbuchLSS,Bruch}
\begin{equation}
I_R(E_s) \propto T_i \, \tau_i \, \left|\left<P_i\right|\mathcal{H}_{EF}(\omega_{ph})\left|{P_s}\right>\right|^2
               \, T_s \, \rho_s \, \delta(E_i-\hbar\omega_{ph}-E_s).
\label{CrossSection}       
\end{equation}
Here $\left|P_i\right>$ and $\left|P_s\right>$ are the initial and final polariton states, 
with energies $E_i$ and $E_s$, respectively. The lifetime of the initial state is $\tau_i$,
while $\rho_s$ represents the final density of states. The scattering process can be described
as follows: a photon of energy $E_i$ impinges on the material's surface and is converted into a
polariton $P_i$ with probability $T_i$. Before dephasing by relaxation processes ocurring in a 
typical time $\tau_i$, the polariton $P_i$ interacts with the ion cores through the electron-phonon
coupling Hamiltonian $H_{EF}$ (photoelastic coupling), creating or anhilating an acoustic phonon of energy $\hbar \omega_{ph}$. 
Finally, the scattered polariton $P_s$ resulting from this interaction leaves the system with a probability 
$T_s$ of converting into a photon of energy $E_s$ at the sample surface.

Let us now discuss this general expression in the specific case of intra-branch lower polariton scattering.
Eq.~\ref{CrossSection} can be expressed in a simpler form using the previously defined photon and exciton strengths:~\cite{SuppInfo} 
\begin{equation}
I_R^{LP} \propto {S_{cav}^{LP}}^{\,2} \, \left({S_H^{LP}}^{\,2} + \alpha_L^2 {S_L^{LP}}^{\,2}\right)
\, \Gamma_{LP}^{-2}.
\label{FinalEquation}
\end{equation}
In this expression we have used $T_{LP} \propto S_{cav}^{LP}$. 
The probability for an external photon to couple with a polariton is proportional to the photon strength of the latter.
$\Gamma_{LP}=\tau_{LP}^{-1}$ is the {\em homogenous} spectral width of the involved polariton state. 
Assigning a lorentzian spectral weight of width
$\Gamma_{LP}$ to the polariton state, given by its lifetime, at resonance $\rho_{LP}$ is
proportional to $\Gamma_{LP}^{-1}$. The polariton lifetime depends on the lifetime of
its components by $\Gamma_{LP}=\Gamma_{cav} S_{cav}^{LP} + \Gamma_H S_H^{LP} + \Gamma_L S_L^{LP}$,
with $\Gamma_{cav}$, $\Gamma_H$ and $\Gamma_L$ the {\em homogenoeus} spectral widths of the cavity
mode, and the two involved exciton states, respectively. To derive Eq.~\ref{FinalEquation} the electron-phonon Hamiltonian matrix element has been expressed as~\cite{SuppInfo} $\left|\left<LP\left|\mathcal{H}_{EF}\right|LP\right>\right|^2 
\propto {S_H^{LP}}^{\,2} + \alpha_L^2 {S_L^{LP}}^{\,2}$ 
with $\alpha_L=\left<E_1L_1\left|\mathcal{H}_{EF}\right|E_1L_1\right>
/\left<E_1H_1\left|\mathcal{H}_{EF}\right|E_1H_1\right>$.
Polaritons indeed interact with the lattice by the deformation potential interaction (characterized by the photoelastic constant) only through their exciton component. 
$\alpha_L$ is determined by the quantum-well exciton wavefunction envelope. Since we are considering here the
fundamental quantum-confined state it should be of the order of 1. 
It is in Eq.~\ref{FinalEquation} that the two central concepts of this work are captured: firstly, the quadratic dependence on $S_X$ expresses the strong double-resonant photoelastic interaction, and secondly a smoothly behaved and narrow $\Gamma_{LP}$ reflects the fact that the LP mode's linewidth (i.e., its Q-factor) is determined by the homogenous broadening of cavity and exciton modes, and is {\em not} significantly altered by losses (e.g., as would appear due to the continuum of e-h excitations). As we will show below, dissipation does indeed determine the polariton mediated optomechanic processes when the higher energy polaritons are involved.

Similar equations can be derived for the other two polariton branches. 
Eq.~\ref{FinalEquation} represents a double resonant 
scattering process, which requires a balance between the photonic and excitonic strengths
of the participating polaritons in order to be maximized. If the photon component dominates, light will couple 
efficiently with the material polaritons, but the latter will couple poorly with the lattice.
The reverse also holds, i.e., a large exciton strength of the polariton assures a
strong interaction with phonons, but a poor coupling efficieny with the outside world.
The optimimum hence should occur when exciton and photon strengths
are similar, that is, close to zero cavity-exciton detuning (see Fig.~\ref{fig02}).
Coming back to Fig.~\ref{fig03}, we show in the top panel the calculated polariton dispersion relationship. Open stars indicate the expected condition for maximum Raman efficiency following the above simplified discussion.That is, maximum efficiency for exciton and photonic strengths perfectly balanced. The maxima of the Raman efficiencies for all three branches fall close to these values, but somewhat shifted towards detunings corresponding to larger excitonic character on each branch. This effect derives from lifetime effects in Eq.~\ref{FinalEquation}, as we discuss next.

\begin{figure} 
    \begin{center}
    \includegraphics[scale=1,angle=0]{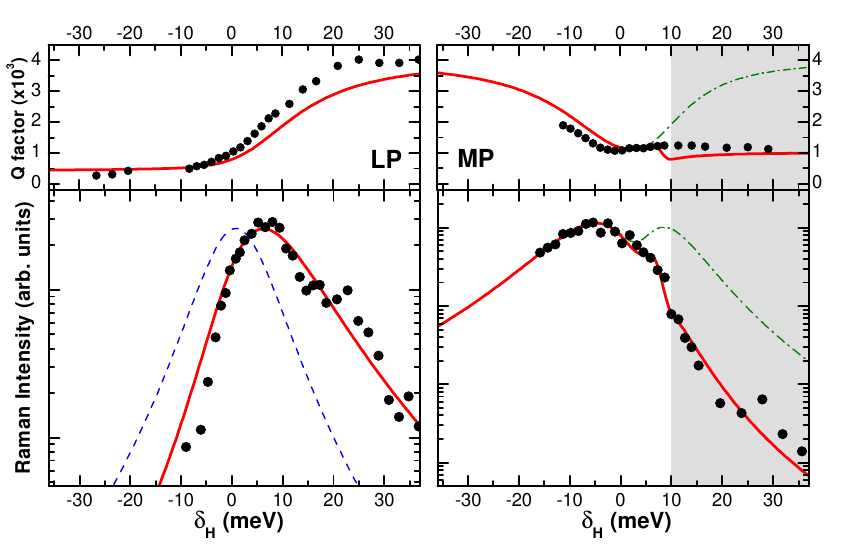}
    \end{center}
    \vspace{-0.8 cm}
\caption{Bottom panels: total Raman integrated intensity as a function of cavity-$E_1H_1$ detuning for the lower (left panel) and middle (right panel) polariton branches. Full red curves are obtained using Eq.~\ref{FinalEquation}. For the LP, the dashed curve corresponds to the calculation without inclusion of lifetime effects. Note the logarithmic scale. Top panels: corresponding cavity polariton mode Q-factor for each branch derived from the fit of the Raman intensities. Solid circles are independent experimental values estimated from the photoluminescence linewidths. Grey regions indicate the continuum of e-h excitations. The dashed-dotted curves in both MP panels are the Raman efficiency and the derived Q-factor calculated using Eq.~\ref{FinalEquation} without inclusion of the dissipative channel related to the continuum of interband e-h transitions.
\label{fig04}}
\end{figure}

In Fig.~\ref{fig04} we present with solid symbols the Raman intensity (integrated over the full Raman spectra) as a function of $\delta_H$ measured for the lower (left) and middle (right panel) polariton branches. Let's first address the LP case. The dashed curve corresponds to a simpler form of Eq.~\ref{FinalEquation} $I_R \propto S_{cav}^i S_X^i S_X^s S_{cav}^s$, i.e., without inclusion of lifetime effects. Besides a rigid red-shift of the curve respect to the experiment, it is quite notable the resemblance, particularly when the {\em only} adjustable parameter in this calculation is the maximum intensity. The blue-shift of the data with respect to the calculation is clearly indicative of lifetime effects. 
In fact, according to Eq.~\ref{FinalEquation} the resonance scan maxima should shift either towards more photonic or excitonic character, depending on whether the cavity confined photon or the exciton, respectively, have a longer lifetime. The continuous curve in Fig.\ref{fig04} (bottom left) corresponds to the full calculation using Eq.~\ref{FinalEquation}. Here we have used the measured linewidth of the cavity mode obtained in the pure photonic regime ($\Gamma_{cav} \sim 3.3$~meV, i.e., a cavity Q-factor~$ \sim 450$), corresponding to a lifetime $\tau_{cav} \sim 0.2$~ps. The only fitting parameters are the maximum Raman intensity, and the ratio $\Gamma_{cav}/\Gamma_H$. The $E_1H_1$-exciton lifetime that is obtained from the fit is $\sim 1.8$~ps (corresponding to $\Gamma_H \sim 0.4$~meV homogeneous linewidth, as compared to $\sim 1.1$~meV inhomogenous broadening measured at very negative detunings), which is reasonable for a GaAs MQW at 80~K. The agreement is remarcable. 
The Q-factor derived from the fit of the Raman intensity using the polariton model is shown with a full line on the top panel in Fig.\ref{fig04} (left). As an additional test for the model, an estimation of the cavity polariton mode Q-factor obtained from the measured photoluminescence linewidths is shown with full circles. Note the contrast with the situation depicted in Fig.~\ref{fig01} for bulk-GaAs. For the strongly coupled cavity structure studied here, with $\Gamma_{cav} > \Gamma_H$, the Q-factor {\em increases} on resonance. These results highlight the main conclusions of this paper, i.e., the presence of a strong double-resonant photoelastic coupling, and the lack of a detrimental influence of the exciton lifetime on the effective polariton Q-factor.

On closing we center our attention on the resonant behavior on the middle polariton, shown on the right panels of Fig.~\ref{fig04}. The dashed-dotted curves correspond to the model using the same fitting parameters as for the LP, taking $\Gamma_L=\Gamma_H$. Note the calculated double maxima, which is due to the separate almost optimum balance of $S_{cav}$ with either $S_H$ or $S_L$. The abrupt decay of the measured Raman intensity around $\delta_H=10$~meV results from the onset of dissipation as the polariton branch overlaps with the continuum of single-particle e-h excitations (indicated with grey background in Figs.~\ref{fig02} and \ref{fig04}), not taken into account by the model. The continuous curve is also obtained using the full calculation based on Eq.~\ref{FinalEquation},  but with such additional interband dissipative mechanism affecting $\Gamma_{MP}$. A good description of the experiment is obtained adding ad-hoc to $\Gamma_{MP}$ a smooth function defined by a step-function of gaussian profile with typical width $\sim 0.5$~meV, approaching zero at large negative detunings, and a constant value $\sim 1.2$~meV for energies larger than the ionization edge (exciton binding energy $\sim 6.5$~meV). 
The MP Q-factor, shown on the top-right panel, strongly drops on the onset of the continuum of e-h excitations (highlighted with the grey shading in the figure). This behavior is conceptually similar to that displayed by the Q-factor of the bulk-GaAs microcavity in Fig.~\ref{fig01}.

The magnitude of the light-sound interaction can be quantified by the optomechanical coupling factor  $g_{om}=d\omega/du$ which describes the variation of the cavity mode angular frequency $\omega$ with the mechanical displacement $u$.~\cite{Ding}  $\Omega_R=g_{om}x_0$ is also used as a measure of the optomechanical coupling. Here $x_0=\sqrt{\hbar/2m_{eff}\Omega_0}$ is the zero-point motion of the mechanical oscillator, and $m_{eff}$ its effective motional mass.  The optomechanical coupling factor in DBR-based GaAs microcavities has been recently calculated to be $g^{ph}_{om}=83$~THz/nm using a value $P=200$ for the photoelastic constant.~\cite{FainsteinPRL2013} Based on the results reported here, assuming that a fully resonant photoelastic coupling can be attained in the strong coupling regime (photoelastic constants in the $5 \times 10^3$ range, see Fig.~\ref{fig01}), we thus estimate enormous optomechanical coupling factors $g_{om}$ in the PetaHz/nm range (PetaHz=10$^{15}$Hz) in {\em polariton} optomechanics. Considering a micropillar $\lambda/2$ cavity of $1~\mu$m radii, for the $\Omega_0=20$~GHz mode $m_{eff} \sim 8$~pg, and thus $\Omega_R \sim 10^9$~Hz (i.e., in the $\sim $~GHz range). We note that other cavity optomechanical realizations typically have $g_{om}$ in the sub-GHz/nm~range, and $\Omega_R$ in the hundreds of kHz.

In conclusion, we have highlighted the problem possed by dissipation to fully exploit the strong resonant disspersive photoelastic mechanism in cavity optomechanics: in principle at the same rythm that the coupling increases, the cavity Q-factor should be destroyed. However, while this is strictly true for bulk-cavities (absorption due to the e-h continuum is equivalent to irreversible loss), and also for MQW embedded structures in the weak-coupling regime (roughness and inhomogeneous broadening play essentially the same role as the continuum in bulk), it is {\em not} the case when cavity polariton optomechanics is considered (strong coupling regime). 
Optomechanical phenomena of a dynamical nature arise when the decay time of the photons inside the cavity is comparable to or longer than the mechanical oscillator period. For the high frequencies of the optomechanical modes in DBR-based semiconductor microcavities ($> 20$~GHz), and the attainable cavity Q-factors (readily above $10^4$) it thus follows that new promissing perspectives that combine the worlds of cavity optomechanics with CQED in the presence of ultra-strong optomechanical coupling can be envisaged.


\end{document}